\documentclass[12pt,a4paper]{article}
\usepackage{graphicx}
\usepackage[cp1251]{inputenc}
\usepackage{amssymb, amsfonts, amsmath}
\usepackage{amscd}
\newtheorem{Def}{Definition}
\newtheorem{thr}{Theorem}
\begin{document}
\title{Separation of variables in Hamilton-Jacobi and Klein–Gordon–Fock equations for a charged test particle in the Stackel spaces of type (1.1)}
\author{V. V. Obukhov}
\date{}
\maketitle
\abstract All equivalence classes for electromagnetic potentials and space-time metrics of Stackel spaces, provided that  Hamilton-Jacobi equation and Klein-Gordon-Fock equation for a charged test particle can be integrated by the method of complete separation of variables are found.  The separation is carried out using the complete sets of mutually-commuting integrals of motion of type (1.1). Whereby in a privileged coordinate system the given equations turn into parabolic type equations. Hence, these metrics can be used as models for describing plane gravitational waves.

\quad


\begin{flushleft}

           \begin{center}
            Tomsk State  Pedagogical University, 60 Kievskaya St., Tomsk, 634041, Russia.

            Tomsk State University of Control Systems and Radioelectronics 40 Lenin Ave., Tomsk, 634050, Russia
           \end{center}
\end{flushleft}
\section{Introduction}	

 The main goal of this work is to complete the classification of Stackel spaces, started in \cite{1} - \cite{3}. In the spaces the Hamilton-Jacobi equation for a charged test particle moving in an external electromagnetic field admits complete separation of variables. The theory of Stackel spaces is presented in \cite{4} - \cite{9}. A detailed bibliography can be found in \cite{10} - \cite{11}.

According to the main theorem proved in \cite{7} - \cite{8}, the Stackel space is characterized by a complete set of geometric objects. The set consists of mutually commuting Killing vector and tensor fields that satisfy some additional conditions, that allow carrying out complete separation of variables in classical and quantum equations of motion of a test particle in the presence of physical fields of different nature. A large number of works have been devoted to the corresponding classification problems. (see, for example, an overview in \cite{2}). This article lists all non-equivalent cases of complete separation of variables of type (1.1).

The classification is carried out for the case when the complete set consists of two Killing tensor fields and one null Killing vector field without any additional restrictions on the electromagnetic field.

\section{Hamilton-Jakobi equation}

Let us present information from the theory of Stackel spaces which will be required further.
Consider a $n$-dimensional Riemannian manifold \quad   $ V_n $, with a coordinate system  \quad $\{ x^i \}$ \quad  ($i,j,k,l= 0,\dots,n-1$) \quad  and a metric tensor, with components \quad $ g_{ij}$ and $g^{ij}$.

    Hereinafter \quad $\delta^{ij},\delta_{ij}, \delta^{i}_{j}$ \quad are Kronecker symbols, \quad $\lambda, \lambda_{i} = const$ \quad (separation constants),
    \quad \quad $\varepsilon, \varepsilon_i$ \quad can have values \quad $0, +1,-1.$\quad

   Functions that depend only on the variable  \quad $x^i $ \quad will be denoted by the lowercase letters with the obligatory lower index \quad $i$,\quad  constants are marked with lower-case letters and a $tilda$ sign. Everywhere in the text, there is a summation within
the specified limits of index change on repeated upper and lower indexes.

Natural Hamilton-Jakobi equation (the term was suggested by Benenti in \cite{12}) has the form:
\begin{align}\label{1}
 H = g^{ij}p_ip_j +2A^i p_i +B = \lambda,\end{align}
where \quad $p_i = \partial S /{\partial x^i} = S_{,i}. $
Particular case of equation \eqref{1} is the regular Hamilton-Jakobi equation:
\begin{align}\label{2}
 H = g^{ij}p_ip_j +2A^i p_i + A_iA^i = \lambda,\end{align}
\begin{Def}
The equation \eqref{1} allows a complete separation of variables if the space \quad $ V_n $ contains a privileged coordinate system \quad $\{ u^i \}$ \quad in which the complete integral can be represented as:
\begin{align}\label{3} S = \sum_{i=0}^{n-1}{s_{i}(u^i)}.\end{align}
In this case, \quad $ V_n $ \quad is called Stackel space.

\end{Def}

\begin{Def}
Let \quad $ V_n $ \quad  has a set of geometric objects, consisting of mutually-commuting $N<n$ Killing vector fields --
\quad $Y_p^{i}\quad $
and \quad  $n-N$ \quad Killing tensor fields -- \quad $X_\nu^{ij}$ \quad (including
metric tensor:
\quad $g^{ij}=X_{n-1}^{ij}),\quad $ satisfying the conditions:

 \quad

 1) From \quad $B^\nu X_\nu^{ij} + B^{pq}Y_p^{i} Y_q^{j}=0$ \quad it follows \quad $B^\nu =
  B^{pq}=0.$

 \quad

 2) There are functions \quad  $B^\nu_{\mu \tau},\quad B^{pq}_{\mu \nu},\quad B^{p}_{\mu q},$ \quad such that

 $$ X_\mu^{il}g_{lk}X_\nu ^{kj}= B^\tau_{\mu \nu} X_\tau ^{ij} +
 B^{pq}_{\mu \nu}Y_p^{i}Y_q^j, $$
 $$  X_\mu^{il}g_{lk}Y_p^{k}= B^{q}_{\mu p}Y_q^i.$$

 3) There are functions, \quad $ B_\nu $ \quad such that  \quad
 $$\det{B_\nu B^\nu_{\mu \tau}}>0.$$ \quad
 In this case the set is called full. Hereinafter \quad $p,q =0, \dots, N-1,
 \quad \nu, \mu, \tau = N, \dots, n-1$.

\end{Def}
Let us denote
$$N_0 = N - rank\|{Y_p}^{i} g_{ij}{Y_q}^j\|.$$
If  \quad $N_0 \ne 0,$\quad the equation \eqref{1}\quad is a parabolic one and among the non-ignored variables \quad $u^\nu$ \quad it can be distinguished as the variables \quad $u^{\nu_0}, \quad (\nu_0, \mu_0 = N, \dots,
 N+N_0) $\quad such that \quad $ g^{\nu_0 \mu_0} = 0.$ \quad It can be shown that if the space has the Lorentzian signature, the number \quad $ N_0 $ \quad can take one of the two values: $0$ or $1$.

\quad

\begin{thr}
%\begin{proof}
The existence of a complete set is a necessary and sufficient condition for the complete separation of variables in the equation \eqref{1}.
%\end{proof}
\end{thr}

\begin{Def}
The space \quad $V_n$,\quad which allows a complete set is called Stackel space of \quad $(n, N, N_0)$\quad type (or simply - of type $(N, N_0)$ for the case of space-time signature).
\end{Def}

\begin{thr}
If the equation \eqref{1} allows a complete separation of variables, the complete set of  the motion integrals can be represented as
\begin{align}\label{4}
 \hat{X}_q  = \delta_{q}^{i} p_i, \quad  \hat{X}_\mu =  (\hat{\phi}^{-1})^\nu_\mu(\varepsilon_\nu p_\nu ^2
+2h_\nu^{\nu q}p_\nu p_q +2h_\nu^q p_q +
 2h_\nu ^\nu p_\nu +h_\nu).
\end{align}
where \quad $ Y_p^i= \delta_{p}^{i}$\quad are Killing vectors. Killing tensors can be represented in the form:
 \begin{align}\label{5}  X_\mu ^{ij} = (\hat{\phi}^{-1})^\nu_\mu(\delta_{\nu}^i\delta_{\nu}^j \varepsilon_\nu  +
\delta_{p}^i\delta_{q}^j h_\nu^{p q } + 2\delta_{p}^i\delta_{\nu}^j h_\nu^{p \nu}),
\end{align}
where  $(\hat{\phi}^{-1})^\nu_\mu$ are elements of the matrix which is inverse of the Stackel matrix  $\|\phi^\mu_\nu\|.$
For the case of the regular equation \eqref{2} the following conditions are satisfied:
\begin{align}\label{6}  A^i = (\hat{\phi}^{-1})^\nu_{n-1} h_\nu^{i},
  \quad  A_iA^i = (\hat{\phi}^{-1})^\nu_{n-1} h_\nu,
\end{align}
\end{thr}
As it was noted above, in the space $ V_4 $ with the Lorentzian signature $ N_0 = 0.1 $. Therefore, there are seven disjoint types of Stackel space-times with the given signature. Respectively: four non-null types: $(0.0), (1.0), (2.0), (3.0)$; and three null types: $(3.1)$, $(2.1)$, $(1.1)$. For the type $(0.0)$ the electromagnetic potential $ A_i = 0 $. In the privileged coordinate systems of space-times of type $ (3.N_0) $ there is only one non ignored variable.  All components of the metric tensors and electromagnetic potentials depend on this variable and Hamilton-Jacobi equation (both and Klein-Gordon-Fock equation) admits complete separation of variables.

The classification of the space-time metrics of type (1.0) was carried out in \cite{1},  the classification of the space-time metrics of type (2.0) - in \cite{2} and the classification of space-time metrics of type (2.1) - in \cite{3}. Thus, it is only left to consider the space-time metrics of type (1.1).

 The metrics of the space-times of type (1.1) have been widely used by researchers in the study of various problems. For example, all privileged coordinate systems and all electromagnetic fields in which the Hamilton-Jacobi both and Klein-Gordon-Fock equations in flat space-time admit complete separation of variables of type (1.1)  were found in the papers \cite{13} - \cite{15}. In the papers \cite{16} - \cite{17} the same  problem was solved for vacuum both and electrovacuum  Einstein equations, as well as in the Vaidya problem: (see \cite{18}-\cite{20}). Separate classification problems were considered for the Einstein equations with dusty matter \cite{21}-\cite{22}.

There is only one Killing vector field in the space-time of type (1.1): \quad $Y^{i}=\delta_0^{i}$.\quad
Thus  \quad $u^0$\quad is the ignored variable. As
$$\quad Y^iY_i=0 \to g_{00}=0 \to g^{11}=0,\quad $$
equation \eqref{1} is parabolic. When considering an electromagnetic field, the presence of a zero non-ignored variable in a privileged coordinate system may mean that this field describes a plane electromagnetic wave. By analogy, some space-time metrics of the type (1.1) can be considered as a model of a plane gravitational wave.
Let us mark the non-ignored variables  with indexes \quad $\alpha,\beta =1, \dots, 3;\quad \nu,\mu = 2,3.$ \quad Whereby $u^1$ is a null variable.

\subsection{Solution of the functional equation \eqref{6}}

Now the metric tensor and Killing tensor can be written in the privileged coordinate system as:
    \begin{align}\label{7} X_1^{ij} = g^{ij}= \frac{1}{\Delta}((\tau_3-\tau_2)h^{ij}_1
    +(\tau_1-\tau_3)h^{ij}_2 +(\tau_2-\tau_1)h^{ij}_3),\cr
      X^{ij}_2= \frac{1}{\Delta}((w_2-w_3)h^{ij}_1 +(w_3-w_1)h^{ij}_2 +(w_1-w_2)h^{ij}_3),
       \cr
       X^{ij}_3= \frac{1}{\Delta}((w_3 \tau_2-w_2 \tau_3)h^{ij}_1 +(\tau_3 w_1-
       \tau_1 w_3)h^{ij}_2 +(\tau_1 w_2-\tau_2 w_1)h^{ij}_3),\end{align}
     where
$$ h_\alpha^{ij}= (\delta^i_0 \delta^j_1 + \delta^j_0 \delta^i_1 )\delta_\alpha ^1 +
\delta^i_\nu \delta^j_\nu \delta_\alpha^\nu \varepsilon_\nu +\delta^i_0 \delta^j_0 b_\alpha.$$
     $$\Delta = (\tau_3-\tau_2)w_1+(\tau_1-\tau_3)w_2 + (\tau_2-\tau_1)w_3,$$
      $$ (\tau_3-\tau_2)=V^1, \quad (\tau_1-\tau_3)=V^2, \quad  (\tau_2-\tau_1)=V^3.$$
	Hence,  the equation \eqref{2} admits a complete separation of variables if the equation \eqref{6}, the components of the metric tensor and electromagnetic potential have the form:
 \begin{align}\label{8} A^\alpha A_\alpha =\frac{(2e_1 a_\alpha -
     e_1^2 b_\alpha)V^\alpha}{\Delta} = \frac{h_\alpha V^\alpha}{\Delta}.\end{align}
   \begin{align}\label{9}(g^{ij})=\begin{pmatrix}\frac{b_\alpha V^\alpha}{\Delta} & \frac{V^1}{\Delta} & 0 & 0\\
   \frac{V^1}{\Delta} & 0  & 0 & 0 \\
    0 & 0 & \frac{\varepsilon^2 V^2}{\Delta} & 0\\ 0 & 0 & 0 & \frac{\varepsilon^3 V^3}{\Delta}\end{pmatrix},\quad
    (g_{ij})=\begin{pmatrix} 0  & \frac{\Delta}{V^1} &
    0 & 0\\ \frac{\Delta}{V^1}  & -\frac{\Delta b_\alpha V^\alpha}{{V^1}^2} & 0 & 0 \\ 0 & 0 & \frac{\varepsilon^2\Delta}{ V^2} & 0
    \\ 0 & 0 & 0 & \frac{\varepsilon^3\Delta}{ V^3}\end{pmatrix},
    \end{align}
\begin{align}\label{10} A^i=\frac{\delta^{i0}(a_\alpha V^\alpha)+\delta^{i1} e_1 V^1}{\Delta},
\quad  A_i= \delta_{i0} e_1+
{\delta_{i1}}\frac{(a_\alpha- e_1 b_\alpha) V^\alpha}{V^1};\end{align}

The functional equation \eqref{8} is reduced to a system of algebraic equations. In order to find all the metric tensors and electromagnetic potentials that allow to realize a complete separation of variables
of type (1.1) in equation (2) it is necessary to solve this system.
Note that using the acceptable transformations of coordinates and gradient transformations of the potential of the form
  $$u^0 \to u^0 + \int{f_1}du^1,\quad A_1 \to A_1+z_1,$$
one can turn the functions $\quad a_1$\quad and \quad $b_1$\quad into zero.  Then the equation (10) will be  represent  in the form:
     \begin{align}\label{11}(2e_1a_2 - e_1^2b_2- h_2 - h_1)(\tau_3-\tau_1)=
     (2e_1a_3 - e_1^2b_3- h_3 - h_1)(\tau_2-\tau_1).\end{align}
	It is equivalent to the system of function equations:
   $$ 2e_1a_\nu - e_1^2b_\nu- h_\nu - h_1 = \theta_1(\tau_\nu - \tau_1).$$
   The system has two solutions.

   $1)$ \quad $\dot{e}_1 \ne 0 \to a_\nu=\tilde{a}t_\nu, \quad b_\nu = \tilde{b}t_\nu,$ \quad which is equivalent to:
   \quad $a_\nu=b_\nu=0,$ \quad where \quad $g^{00}=A^0=A_1=0.$

 $2)$ \quad $e_1=\tilde{e}$,\quad which is equivalent to:\quad $e_1=0.$ \quad
 For both cases we have:
  $$A_i A^i=0\quad \to h_\alpha=0.$$

\qquad
\subsection{Results}

These solutions can be united. For this purpose let us introduce the symbols \quad $\xi_0, \xi_1$ \quad satisfying the following conditions:
$$
 \xi_0\xi_1=0, \quad \xi_0 + \xi_1=1.
$$
Then the metric tensor and the electromagnetic potential will take the form:
   \begin{align}\label{12}(g^{ij})=\begin{pmatrix}\xi_1 \frac{b_\nu V^\nu}{\Delta} & \frac{V^1}{\Delta} & 0 & 0\\
   \frac{V^1}{\Delta} & 0  & 0 & 0 \\
    0 & 0 & \frac{\varepsilon^2 V^2}{\Delta} & 0\\ 0 & 0 & 0 & \frac{\varepsilon^3 V^3}{\Delta}\end{pmatrix},\quad (g_{ij})=\begin{pmatrix} 0  & \frac{\Delta}{V^1} &
    0 & 0\\ \frac{\Delta}{V^1}  & -\xi_1\frac{\Delta b_\nu V^\nu}{{V^1}^2} & 0 & 0 \\ 0 & 0 & \frac{\varepsilon^2\Delta}{ V^2} & 0
    \\ 0 & 0 & 0 & \frac{\varepsilon^3\Delta}{ V^3}\end{pmatrix},
    \end{align}

    $$ A^i=\frac{\delta^{i0}\xi_1(a_\nu V^\nu)+\delta^{i1}\xi_0 e_1 V^1}{\Delta},\quad  A_i= \delta_{i0}\xi_0 e_1+
    \frac{\delta_{i1}\xi_1(a_\nu V^\nu)}{V^1};
    $$

  Functions  $s_\nu$  from the equation \eqref{3}  have the form:

 $$ s_1 = \int\frac{\lambda_1+\lambda_2 \tau_1+\lambda_3 w_1}{2(\lambda_0 +\xi_0 e_1)}du^2,$$
$$
s_\nu = {\varepsilon} \int\sqrt{\varepsilon^\nu[(\lambda_1+ \lambda_2 \tau_\nu + \lambda_3 w_\nu)- \xi_1( b_\nu {\lambda_0}^2 + 2a_\nu \lambda_0 )]}du^3
$$

\section{Separation of variables in linear differential equations of the second order}

Let us consider the linear differential equation
\begin{align}\label{13} \hat{H}\Phi=(g^{ij}\hat{\partial}_i\hat{\partial}_j +
W^i\hat{\partial}_i + W)\Phi=\lambda\Phi,\end{align}
where \quad $\Phi, \quad W,\quad W^i\quad $ are complex functions.
The theory of variables separation in equation \eqref{13} was constructed in the papers \cite{23} - \cite{24}. Let us present the data from this theory which will be necessary further.

\qquad

\begin{Def}

\quad

Operators $ \hat{X} $ and $ \hat{X}' $  are  equivalent according to the function F, if:
  $$ \hat{X}' = \exp{(-F)}\hat{X}\exp{F}. $$
\end{Def}

\begin{Def}

\qquad

   The equation \eqref{13} admits a complete separation of variables provided there are function
    \quad $F$\quad and the privileged coordinate system
     \quad $ \{ u^i \} $ \quad that allow to represent the solution of the equation
 \quad$$ \hat{H}'\phi = \lambda\phi$$
into the following form:
 %\quad (\hat{X}_{n-1}=\hat{\textrm{H}}, \quad \lambda_{n-1} = \lambda)
   \begin{align}\label{14}
  \phi =\prod_{i}\varphi_i (u^i\lambda_j),\quad \det||\frac{\partial^2 {\phi}}{\partial{u^i}\partial{\lambda_j}}|| \ne 0 ;
  \end{align}
where \quad $\lambda_i$ \quad essential parameters \quad $(\lambda_{n-1}=\lambda)$.
%$\lambda_j $ eigenvalues of reciprocal-commuting symmetry operators
%$ \hat{X}'_i$:
%
%$$\hat{X}'_i\phi=\lambda_i\phi.$$
%Здесь
%$$ \hat{X}_{n-1}=\hat{H},\quad \lambda_{n-1}=\lambda,\quad [\hat{X}_i ,\hat{X}_j]=0. $$
The solution to the equation \eqref{13} is represented with \quad $\phi$\quad and\quad $F$\quad as follows:

 $$\Phi =\exp{F(u)}\phi$$

\end{Def}

 \begin{thr}

\quad

The equation \eqref{13} admits the complete separation of variables if and only if there is a complete set of mutionally-commuting symmetry operators.
 \begin{align}\label{15} \hat{X}_p=Y_p ^i\hat{\partial}_i, \quad \hat{X}_\nu = {X_\nu}^{ij}\hat{\partial}{_i} \hat{\partial}{_j} +
  2{X_\nu}^{i}\hat{\partial}{_i} +X_\nu,\quad ([\hat{X}_i ,\hat{X}_j]=0,
  \quad\hat{X}_{n-1} =\hat{H}).\end{align}
The essential parameters \quad $\lambda_i$\quad introduced above are eigenvalues, and the function $\Phi$ is the eigenfunction of the operators $\hat{X}_i$:
  $$\hat{X}_i\Phi=\lambda_i\Phi.$$
   \end{thr}
 Obviously, the functions \quad  ${X_\nu}^{ij}, \quad {Y_p}^{i} $ \quad  form a complete set in according to the Definition 2.

	Note that the condition of mutually commutativity  of the symmetry operators of the equation \eqref{13} is not necessary for  the exact integration, since the theory of Stackel spaces can be generalized for a case of non-commutative operators of motion (see, \cite{25}  \cite{26}).

\begin{thr}

\qquad
 If the equation \eqref{13} admits complete separation of variables in the privileged coordinate system $ \{ u^l \} $ in the same coordinate system the equation \eqref{1} admits complete separation of variables too. Thus the $V_n$ belongs to the Stackel space

\end{thr}

A theorem similar to Theorem 2 is valid:

\begin{thr}

\qquad

If the equation \eqref{13} admits a complete separation of variables the complete set of the symmetry operators \eqref{15} can be presented as:
 \begin{align}\label{16}
 \hat{X}_q  = \hat{\partial}_q, \quad  \hat{X}'_\mu = (\hat{\phi}^{-1})^\nu_\mu \hat{H}_\nu.
\end{align}
where
$$
 \hat{H}_\nu =
 \varepsilon_\nu\hat{\partial} _\nu ^2 + h^{pq}_\nu \hat{\partial}_p \hat{\partial}_q +
 2h_\nu^{\nu q}\hat{\partial}_\nu \hat{\partial}_q +2h_\nu^q \hat{\partial}_q +
 2h_\nu ^\nu \hat{\partial}_\nu +h_\nu.
$$
\end{thr}
	
Klein-Gordon-Fock equation is a particular case of the equation \eqref{13}:
 \begin{align}\label{17}\hat{H}\Phi = (-\imath\hat{\nabla}^i +A^i)(-\imath\hat{\nabla}_i +A_i)\Phi=\lambda\Phi,\end{align}
 Using the metrics of Stackel spaces one can represent the equation \eqref{17} in the form:
 \begin{align}\label{18}
 [G^{ij}\hat{\partial}_i \hat{\partial}_j+(G^{ij}_{,i}+G^{ij}\chi_{,i}+2\imath B^{j})\hat{\partial}_{,j}+
\imath(B^i_{,i}+\chi_{,i} B^i)-
B^iA_i +\lambda\Delta]\Phi=0,
 \end{align}
 where:
 $$ G^{ij}=g^{ij}\Delta, \quad B^i = A^i\Delta, \quad  G=det\|G^{ij}\|, \quad \chi = \frac{1}{2}(\ln{\frac{\Delta^{n-2}}{G}}),
  \quad \Delta=\det\|{\varphi}^\nu _\mu\|. $$

  %We will consider equation \eqref{18} under the condition that the Hamilton-Jacobi equation for a charged particle admits the separation of variables. Then the relations \eqref{6} are valid.
%
%  The sufficient condition of the complete separation of variables in the equation \eqref{18} has the following form:
We will consider the equation \eqref {18} provided that the Hamilton-Jacobi equation for a charged particle admits complete separation of variables. Then the relations \eqref {6} are true. In addition, we assume that the following condition is satisfied:
\begin{align}\label{21} \chi = \ln{\alpha_1 \alpha_2 \alpha_3} \quad \to \Delta^2 = G{\alpha_1 \alpha_2 \alpha_3}.\end{align}
In this case the function  $F=1$, and equation \eqref{18} admits complete separation of variables.

\subsection{Solution of the functional equation}

For the space-time of type $(1.1)$ with metric tensor \eqref{12} the operator \quad $\hat{H}$ \quad can be written in the form:
\begin{align}\label{19} \hat{H}=\frac{V^\alpha \hat{H}_\alpha
}{\Delta},\end{align}
where
$$\hat{H}_1 = 2(\hat{\partial}_0 +\imath \xi e_1)\hat{\partial}_1  + \dot{\alpha}_{1}(\hat{\partial}_0 + \imath \xi_0 e_1)
+\imath\xi_0\dot{e}_1,$$
$$ \hat{H}_\nu =\varepsilon_\nu(\hat{\partial}_\nu^2 +
 \dot{\alpha}_{\nu}\hat{\partial}_\nu) +
\xi_1( b_\nu\hat{\partial}_0^2+2\imath a_\nu \hat{\partial}_0).
$$
Symmetry operators have the form:
\begin{align}\label{20}
\hat{X}_1=\frac{(\tau_2\omega_3-\tau_3\omega_2)\hat{H}_1+(\tau_3\omega_1-\tau_1\omega_3)\hat{H}_2 + (\tau_1\omega_2-\tau_2\omega_1)\hat{H}_3}{\Delta},
 \cr
\hat{X}_2 = \frac{(\omega_2 - \omega_3)H_1 + (\omega_3 - \omega_1)H_2 + (\omega_1 - \omega_2)H_3}{\Delta},\quad \hat{X}_0=\delta_0^i\hat{\partial}_i.
\end{align}

%The sufficient condition of the complete separation of variables in the equation \eqref{18} has the following form:
%\begin{align}\label{21} \chi = \frac{1}{2}(\ln{\frac{\Delta^2}{{V^1}^2 V^2 V^3}}) = \ln{\alpha_1 \alpha_2 \alpha_3} \quad \to \frac{\Delta}{\alpha_1 \alpha_2 \alpha_3(\tau_2-\tau_3)}  =  \sqrt{\varepsilon (\tau_2-\tau_1)(\tau_3-\tau_1)}.\end{align}

Then the separated system can be represented as:
\begin{align}\label{23}\imath[2(\lambda_0+\xi_0 e_1)\hat{\partial}_1 + \dot{\alpha_1}(\lambda_0 - e_1\xi_0)
  +\xi_0\dot{e}_1]\varphi_1  =
  (\lambda_1 + \lambda_2\tau_1 + \lambda_3\omega_1)\varphi_1, \cr
   [\varepsilon_\nu(\hat{\partial}_\nu^2 +\dot{\alpha}_\nu \hat{\partial}_\nu) - \xi_1(b_\nu {\lambda_0}^2 +
  2 a_\nu \lambda_0 )]\varphi_\nu = (\lambda_1 + \lambda_2\tau_\nu+ \lambda_3\omega_\nu)\varphi_\nu.\end{align}
The functional equation \eqref{21} has the form:
$$\label{22} \frac{\Delta}{\alpha_1 \alpha_2 \alpha_3(\tau_2-\tau_3)}=\sqrt{\varepsilon (\tau_2-\tau_1)(\tau_3-\tau_1)}.$$
 The left side of the equation  is a rational function containing functions ($\alpha_\alpha,\tau_\alpha, w_\alpha $), each of them depends on one variable. The right side is an irrational function. Therefore the equation  has solutions provided that
  \quad $$\dot{\tau}_1(\dot{\tau}_2 ^2+\dot{\tau}_3 ^2)=0. $$ \quad
Let us consider the function equation \eqref{22} for all options.

  \quad

   $\mathbf{I}.$ $\dot{\tau}_1=0 $.
  The relations \eqref {12} can be presented as follows:
  $$\hat{g}^{ij}=\begin{pmatrix}\xi_1\frac{b_2 +b_3}{\Delta} & \frac{\tau_2+\tau_3}{\Delta} &
    0 & 0\\\frac{\tau_2+\tau_3}{\Delta} & 0  & 0 & 0 \\ 0 & 0 & \frac{1}{\Delta} & 0\\ 0 & 0 & 0 & \frac{1}{\Delta}\end{pmatrix}, \quad  A^i=\frac{\xi_1 \delta^{i0}(a_2 + a_3)+\xi_0 \delta^{i1}e_1 (\tau_2+\tau_3)}{\Delta},$$
    $$ \Delta = w_1(\tau_2+\tau_3)+w_2+w_3.$$
It allows to represent the functional equation in the form:
\begin{align}\label{23} w_1 (\tau_2+\tau_3) + w_2 + w_3 = {\alpha_1 \alpha_2 \alpha_3 (\tau_2+\tau_3)} .\end{align}
Let us consider possible options.

$\mathbf{a}.$ \quad $\dot{\alpha}_1 = 0$,\quad which is equivalent to \quad $ w_1 = 0.$
\quad Solution of the equation \eqref{23} has the form:
$$ \Delta =\frac{(\tau_2+\tau_3)}{\tau_2 \tau_3 }\quad (\alpha_1=1, \quad \alpha_\nu=\frac{1}{\tau_\nu}).$$

$\mathbf{b}.$ \quad $\dot{\alpha}_1 \ne 0.  \quad  \to  w_1 =
\alpha_1 $, \quad  and from the equation \eqref{23} it follows:
$$ \alpha_2 \alpha_3 =1, \quad  w_\nu = 0 \quad \to \quad \Delta =
\alpha_1(\tau_2 +\tau_3)$$

$\mathbf{II}.$ \quad $\dot{\tau}_1 \ne 0 \to \tau_\nu = (-1)^\nu, \quad \dot{w}_2 \dot{w}_3 = 0.$ \quad  The relations \eqref {12} can be presented as follows:

$$\hat{g}^{ij}=\begin{pmatrix}\xi_1\frac{b_2(\tau_1-1) +b_3(\tau_1+1)}{\Delta} & \frac{1}{\Delta} &
    0 & 0\\\frac{1}{\Delta} & 0  & 0 & 0 \\ 0 & 0 & \frac{\varepsilon_2(\tau_1-1)}{\Delta} & 0\\ 0 & 0 & 0 & \frac{\varepsilon_3(\tau_1 +1)}{\Delta}\end{pmatrix}, $$
    $$ A^i=\frac{\xi_1 \delta^{i0}(a_2(\tau_1-1) + a_3(\tau_1+1))+\xi_0 \delta^{i1}e_1}{\Delta}, \quad \Delta = w_1+w_2(\tau_1-1)+w_3(\tau_1+1).$$

The following options are possible:

\quad

$\mathbf{a}.$ \quad $ w_3 = 0, \quad \dot{w}_2 \ne 0.$ \quad The solution to the equation \eqref{22} has the form:
$$\Delta = (\tau_1 - 1)\exp{\alpha_2}.$$

 $\mathbf{b}.$ \quad $ w_\nu = 0 $ \quad
From \eqref{23} it follows:
 $$\Delta = \alpha_1.$$

\quad
\subsection{Results}

\quad

Let us list all sets containing the obtained metrics and electro-magnetic potentials, for which the Kleyn-Gordon-Fock equation \eqref{18} admits the complete separation of variables.  The function \quad
$$ \Phi= \varphi_1 \varphi_2\varphi_3\exp{\imath \lambda_0 u^0},$$
is an eigenfunction of the symmetry operators  $\quad \hat{X}_i$ \quad  with eigenvalues \quad $\lambda_i$:
$$\hat{X}_i\Phi=\lambda_i. \Phi$$

\quad

$\bf{Set 1}.$

\quad

Metric:
\begin{align}\label{24}
 ds^2 = (2du^0 du^1-{du^0}^2 \xi_1 \frac{b_2 + b_3}{\tau_2 + \tau_3} + (\tau_2 +
  \tau_3 )({du^2}^2 + {du^3}^2))\exp{\beta_1}.\end {align}
  Electro-magnetic potential:
  $$ A_i=\xi_1 \delta_{i1}\frac{a_2 + a_3}{\tau_2 + \tau_3}+\xi_0 \delta_{i0}e_1.$$

  \quad

  \bf{Set 2}.

 \quad

 Metric:
 \begin{align}\label{25}
 ds^2 = \frac{1}{\tau_2 \tau_3}(2du^0 du^1 - {du^1}^2 \xi_1 \frac{b_2 + b_3}{\tau_2 +
  \tau_3} + (\tau_2 + \tau_3 )({du^2}^2 + {du^3}^2)).\end {align}
 Electro-magnetic potential:
  $$A_i=\xi_1 \delta_{i1}\frac{a_2 + a_3}{\tau_2 + \tau_3}+\xi_0 \delta_{i0}e_1.$$

 \bf{Set 3.}

 \quad

 Metric:
\begin{align}\label{26}
 ds^2 = 2du^0 du^1 - {du^1}^2 \xi_1 (b_2 \exp{\beta_1} + b_3 \exp{\gamma_1}) +
 {du^2}^2 \exp{-\beta_1} +\cr {du^3}^2 \exp{-\gamma_1}.\end {align}
 Electro-magnetic potential:
  $$ A_i=\xi_0 \delta_{i0}e_1 + \xi_1 \delta_{i1}(a_2 \exp{\beta_1} +
 a_3 \exp{\gamma_1}).$$

\bf{Set 4.}

 \quad

  Metric:
\begin{align}\label{27}
 ds^2 =( 2du^0 du^1 - {du^1}^2 \xi_1 (b_2  + b_3 \exp{\gamma_1}) + {du^2}^2 + {du^3}^2 \exp{-\gamma_1})\exp{\beta_2}.\end {align}
 Electro-magnetic potential:
   $$A_i=\xi_0 \delta_{i0}e_1 + \xi_1 \delta_{i1}(a_2  + a_3 \exp{\gamma_1}).$$

\section{Conclusion}

The main result of this work is the final solution to the classification problem considered in the papers \cite{1}-\cite{3}.
Thus all Stackel space-time metrics and electro-magnetic potentials of the external electromagnetic field for which the Hamilton-Jacobi equation admits complete separation of variables were found.
 From a physical point of view, attention to external fields is due to the presence of physical processes, in the study of which the influence on the geometry of all fields except the gravitational one can be neglected. As an example, we will present axion fields, which are currently considered as the closest candidates for describing dark matter. When constructing cosmological models of the early Universe, they make a significant contribution to the energy-momentum tensor and therefore cannot be considered as external. \cite{27}-\cite{29}. At the same time, when studying a set of local effects in strong gravitational fields, the electromagnetic and axion fields, due to their insignificant contribution to the formation of space-time geometry, often need to be considered just as external against the background of a given metric. Thus, in \cite{30}, the axion, and electromagnetic fields are related by the following system of equations:
\begin{align}\label{50}\nabla_i(F^{ij} + \Phi F^{*ij})=0,
\quad g^{ij}\nabla_i\nabla_j \Phi +\frac{1}{2}\frac{\partial V(\Phi)}{\partial \Phi} = -\tilde{\kappa}^2 F^{*ij}F_{ij},\end{align}
where \quad  $F^{*ij}$ \quad is the electromagnetic field tensor dual to \quad $ F_ {ij} $ \quad, \quad $g^{ij}$ \quad is a given metric tensor. In some local problems, the function \quad $V$ \quad  can be considered null and the second equation of the system is a linear inhomogeneous equation with respect to the function $\Phi$. The problem of classifying the Stackel metrics, electromagnetic potentials, and the scalar field satisfying the system of equations \eqref{50}
seems to be quite interesting from the physical point of view.

%See also papers \cite{Axi5}-\cite{Axi6}.

We note two more promising areas of application of Stackel spaces theory in the theory of gravity.

1) In multidimensional theories of gravity. The interest in multidimensional spaces is due to the emergence of generalizations of the Kaluza-Klein theory, as well as superstring theory.
Large number of papers have been devoted to the problems of multidimensional gravity (see, for example, \cite{31} - \cite{34}).

2) As already noted, among the Stackel space-times of type $(N.1)$ there are wave-like ones (see article \cite{35} too). Recently, interest in wave metrics has increased significantly. For example, in the papers \cite{36}-\cite{37} propagation of gravitational waves in accelerating universe was described. The results obtained in our paper allow us to construct realistic models with gravitational and electromagnetic plane waves.

These tasks are supposed to be considered in the future.

\quad

Acknowledgment.

This work was supported by the Ministry of Science and Higher Education of the
Russian Federation, project FEWF-2020-0003.

%\section{{References}}

\end{document}